\documentclass[aps,prl,twocolumn,showpacs,preprintnumbers,amsmath,amssymb]{revtex4}

\usepackage{bm}

\usepackage{epsfig,amsopn}

\usepackage{graphicx}

\usepackage{amsmath,amssymb}

\usepackage{natbib}

\renewcommand{\section}[1]{{\par\it #1.---}}

\newcommand{\beq}{\begin{equation}}

\newcommand{\eeq}{\end{equation}}

\newcommand{\bes}{\begin{subequations}}

\newcommand{\ees}{\end{subequations}}

\newcommand{\bea}{\begin{eqnarray}}

\newcommand{\eea}{\end{eqnarray}}

\newcommand{\ba}{\begin{array}}

\newcommand{\ea}{\end{array}}

\newcommand{\beqn}{\begin{eqnarray*}}

\newcommand{\eeqn}{\end{eqnarray*}}

\newcommand{\f}[2]{\frac{#1}{#2}}

\newcommand{\g}{\gamma}

\newcommand{\n}{\eta}

\newcommand{\la}{\langle}

\newcommand{\ra}{\rangle}

\def\nn{\nonumber}

\begin{document}

\title{Crossover from Fermi-Pasta-Ulam to normal diffusive behaviour in heat conduction through open anharmonic lattices}

\author{Dibyendu Roy}

\affiliation{Department of Physics, University of Cincinnati, Cincinnati, Ohio 45221, USA}

\date{\today}

\begin{abstract}
We study heat conduction in one, two and three dimensional anharmonic lattices connected to stochastic Langevin heat baths. The inter-atomic potential of the lattices is double-well type, i.e., $V_{\rm DW}(x)=k_2x^2/2+k_4 x^4/4$ with $k_2<0$ and $k_4>0$. 
We observe two different temperature regimes of transport: a high-temperature regime where asymptotic length dependence of nonequilibrium steady state heat current is similar to the well-known Fermi-Pasta-Ulam lattices with an inter-atomic potential, $V_{\rm FPU}(x)=k_2x^2/2+k_4 x^4/4$ with $k_2,k_4>0$. A low temperature regime where heat conduction is diffusive normal satisfying Fourier's law. We present our simulation results at different temperature regimes in all dimensions. 
\end{abstract}      


\maketitle
Search for a microscopic lattice Hamiltonian showing diffusive normal heat conduction satisfying Fourier's law has been an open long-standing problem \cite{Bonetto00, LLP03}. It is now widely accepted that heat conduction in one-dimensional (1D) momentum-conserving lattice models is anomalous and does not obey Fourier's law, $\bar{J}(\bar{x})=-\kappa \bar{\nabla} T(\bar{x})$, relating the local heat current density $\bar{J}(\bar{x})$ at a point $\bar{x}$ to the local temperature $T(\bar{x})$ gradient ($\kappa$ defines the thermal conductivity which is expected to be an intrinsic property of the material) \cite{Dhar08}.
An exception is a chain of coupled rotators exhibiting normal transport properties even in the absence of an on-site potential (i.e., momentum conserving lattice)\cite{Giardina00, Gendelman00}. This model shows a transition from infinite to normal thermal conductivity with the increase of the local temperature. The typical Hamiltonian of a homogeneous 1D momentum conserving lattice with inter-site potential $V(x)$ is
\bea
H &=& \sum_{l=1}^N \f{1}{2}  \dot{x}_l^2  + \sum_{l=1}^{N+1} V(x_l-x_{l-1}),\label{Ham}
\eea
where $x_l$ is a scalar displacement at lattice site $l$. We use fixed boundary, i.e., $x_0=x_{N+1}=0$. The inter-site potential for a chain of coupled rotators is given by $V(x)=1-\cos(x)$ where $x$ is relative rotation between two neighboring sites of the chain. A variant of the coupled rotator model, namely a double-well potential $V_{\rm DW}(x)=k_2x^2/2+k_4x^4/4$ with $k_2<0$ and $k_4>0$ has been also studied in Ref.\cite{Giardina00}. It is claimed that heat conduction in this model shows normal behaviour at low temperature. In the first part of this paper we study the double-well (DW) potential model in 1D rigorously at low and high temperatures using simulation with classic Langevin model of heat baths.  We observe two different temperature regimes: a high-temperature regime where asymptotic length $(N)$ dependence of thermal conductivity is $\kappa \sim N^{1/3}$ in 1D. This asymptotic length dependence is well-known for 1D Fermi-Pasta-Ulam (FPU) lattices with an inter-site potential $V_{\rm FPU}(x)=k_2x^2/2+k_4x^4/4$ where $k_2,k_4>0$ \cite{Narayan02}. One interesting feature of the DW model is that we find the asymptotic length dependence of $\kappa$ at high temperature for relatively much smaller length (almost two order of magnitude) compared to the length scale where it has been shown for the FPU chains \cite{Mai07}. This occurs because of the absence of low energy long wavelength heat carrying modes in the DW model. We could not unambiguously confirm normal heat conduction at low temperature in this model. We present detailed simulation results in this regime and give several arguments supporting our results.

\begin{figure}[t]
\includegraphics[width=2.5cm]{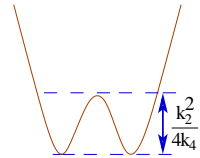}
\caption{A schematic of the double-well inter-particle potential showing height of the potential barrier $k_{\rm B}\bar{T}_1=k_2^2/4k_4$.} 
\label{DW}
\end{figure}

Next we consider lattices with DW type inter-atomic potential in higher dimensions. From our results for 1D we expect the asymptotic length dependence for higher dimensions at high temperature to be similar to that of the FPU lattices. It has been predicted that $\kappa \sim {\rm log}(N)$ for a two dimensional (2D) FPU lattice where $N$ is length along the applied temperature difference \cite{Lepri98, Narayan02}. Although  available numerical simulations for 2D FPU system show $\kappa \sim N^{0.22}$ \cite{Lippi00,Grassberger02, Saito10}. For a three dimensional (3D) FPU lattice, based on kinetic theory and Boltzmann equation approaches, the expectation is that the thermal conductivity should be finite (and Fourier's law is valid) at high temperature where Umklapp processes dominate. 
  Recent works \cite{Saito10, Wang10} claimed to see evidence of normal heat conduction  for the 3D FPU lattices even in the absence of on-site pinning potential. We find in our simulation $\kappa \sim {\rm log}(N)$ for the 2D DW model at high temperature. Thus we give first confirmation of the predicted logarithmic behaviour for the 2D FPU lattices from our simulation.  The high-temperature heat conduction in the 3D DW model does show signature of normal diffusive behaviour. 
 We also simulate low-temperature thermal transport for the DW model in higher dimensions, and we find normal diffusive heat conduction showing a finite $\kappa$ for 2D and 3D lattices. 

The DW model was studied in Ref.\cite{Giardina00} using Nose-Hoover model of thermostat which does not ensure correct energy distribution for the system \cite{Fillipov98}. We here use stochastic Langevin heat baths and ensure local thermal equilibration in our simulation. The DW model experiences a structural phase transition with the reduction of its temperature, and it shows up in different asymptotic length dependence of heat current at low and high temperatures. We can have a rough estimate of the crossover temperature from the following arguments. The height of potential barrier between the stable minima and the unstable maxima is given by $\Delta U=k_2^2/4k_4$ (see Fig.\ref{DW}). Comparing $\Delta U$ with the mean temperature of the baths we get a crossover temperature $\bar{T}_1$ scale for the 1D DW model, i.e., $k_{\rm B}\bar{T}_1=k_2^2/4k_4$ with $\bar{T}=(T_{\rm L}+T_{\rm R})/2$. In our simulation we connect the two end sites of the chain described by Eq.(\ref{Ham}) along with the DW form of the inter-site potential to two Langevin heat baths which are kept at temperatures $T_{\rm L}$ and $T_{\rm R}$. The bath $l$ introduces a fluctuating zero-mean Gaussian white noise $\n_l$ and a dissipative term $\g_l$ in the equation of motion of the end particle of the chain.  The terms $\g_l, \eta_l$ are related by fluctuation-dissipation theorem as given in Eq.\ref{FD}. The coupling strength to the bath is controlled by the dissipation constant $\g_l$.    
The equations of motion for the sites on the wire are:
\bea
\ddot{x}_l&=&-k_2(2 x_l -x_{l-1}-x_{l+1})+k_4(x_{l+1}-x_l)^3\nn\\&-&k_4(x_{l}-x_{l-1})^3+(-\gamma_{\rm L} \dot{x}_1 +\n_{\rm L})\delta_{l,1}\nn\\&+&(-\gamma_{\rm R} \dot{x}_N +\n_{\rm R})\delta_{l,N},\label{LE}
\eea
where the noise-noise correlation is given by
\bea
\la \eta_l(t) \eta_m(t') \ra &=&2\gamma_l k_{\rm B}T_l \delta_{lm}\delta(t-t')~.\label{FD}
\eea
We set the Boltzmann constant $k_{\rm B}=1$. We simulate the above set of equations of motion using a velocity-Verlet algorithm.  We measure the heat current and local temperature in the nonequilibrium steady state (after $10^8$ to $10^9$ time steps of transient dynamics) using following definitions. We define time-averaged local heat current at site $l$ as
\bea
J^l&=&\la j_l(t)\ra_t \equiv {\rm lim}_{t\to \infty} \f{1}{t}\int_0^t j_l(\tau)d\tau \label{HC}\\
{\rm where}~~j_l(\tau)&=&-\dot{x}_l(\tau)\partial_{x_l}V(x_{l+1}(\tau)-x_l(\tau))~~\nn
\eea
\begin{figure}[t]
\includegraphics[width=8.0cm]{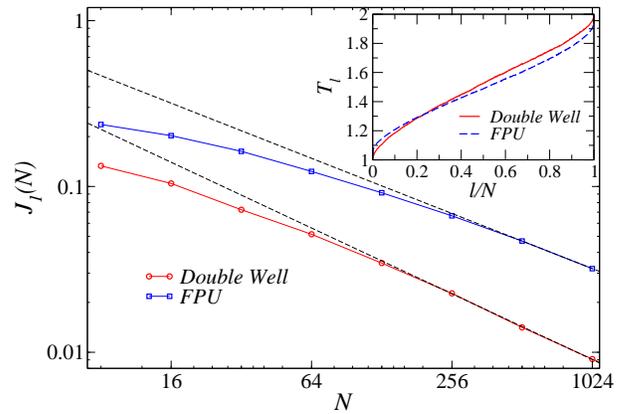}
\caption{Plot of $J_1(N)$ with $N$ for the 1D DW model with $k_2=-1$, $k_4=1$ and for the FPU chain  with $k_2=k_4=1$ (Length up to 1024) at high temperatures, $T_{\rm L}=1$ and $T_{\rm R}=2$. The dotted straight lines correspond to the asymptotic length dependence of $J_1(N)$ for the two models at this length. The inset shows local temperature profiles of the 1D DW and FPU models for $N=1024$. In all plots, $\g_{\rm L}=\g_{\rm R}=1.$} 
\label{FPU}
\end{figure}
We take average over all local heat current to find the global heat current, $J_1(N)=\sum_{l=1}^{N-1} J^l/(N-1)$. The steady state local temperature is given by $k_{\rm B}T_l=\la \dot{x}_l(t)^2\ra_t$. First we simulate the model for temperature of the heat baths being large compared to $\bar{T}_1$.  For example, we choose temperature of the baths are, $T_{\rm L}=1$ and $T_{\rm R}=2$ for $k_2=-1, k_4=1$ and $\bar{T}_1=0.25$. We find that the asymptotic length dependence of heat current for the 1D DW model at high temperatures is similar to the FPU chain which is $\kappa \sim N^{1/3}$. We plot length dependence of $J_1(N)$ for the DW model in  Fig.\ref{FPU}, and compare it to that of the FPU chain for similar length scales. We find $J_1(N)$ falls relatively slower with increasing length for the FPU chain compared to the 1D DW model for the simulated length here. The  asymptotic behaviour $\kappa \sim N^{1/3}$ in FPU chain was shown for a chain length of order of $10^5$ \cite{Mai07}. The local temperature $T_l$ of both the models has been included in the inset of Fig.\ref{FPU} for $N=1024$. It also shows better local thermalization for the 1D DW model than the FPU chain for shorter lengths.

\begin{figure}[htb]
\includegraphics[width=8.0cm]{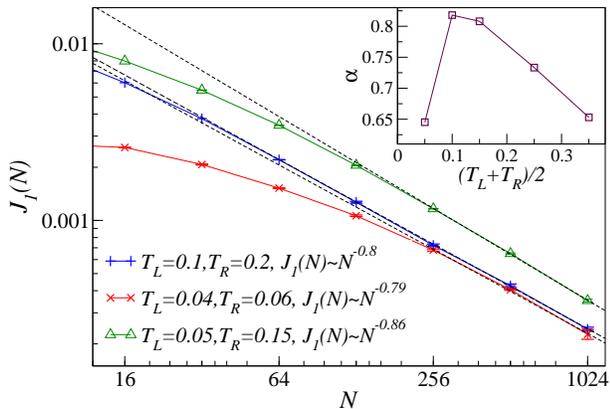}
\caption{Plot of $J_1(N)$ with $N$ for the 1D DW model with $k_2=-1$, $k_4=1$. The dotted straight lines correspond to the asymptotic length dependence of $J_1(N)$ at different mean temperatures. The inset shows $\alpha={\rm Log}[J(128)/J(256)]/{\rm Log}[2]$ with $\bar{T}=(T_{\rm L}+T_{\rm R})/2$ where $(T_{\rm R}-T_{\rm L})<\bar{T}$. In all plots, $\g_{\rm L}=\g_{\rm R}=1.$} 
\label{DWL}
\end{figure}
Now we reduce the mean temperature of the heat baths below $\bar{T}_1$. The steady state heat current at low temperature in the DW model shows faster decay with length compared to high temperature. It has been claimed in Ref.\cite{Giardina00} on the basis of Green-Kubo formula that the low temperature thermal conductivity is independent of $N$ for Nose-Hoover heat baths. It is worthy to mention  that the uncertainty in the simulated thermal conductivity for the DW model in Fig.1 of Ref.\cite{Giardina00} is quite large. Therefore it is hard to conclude from their results that the low temperature heat conduction is normal in the 1D DW model. We simulate the 1D DW model here with Langevin heat baths at different mean temperatures (see Fig.\ref{DWL}). We find that it is difficult to simulate longer chain at lower temperatures keeping the numerical errors in simulated heat current relatively small. Also the length dependence of heat current varies non-monotonically with mean temperature for a finite length as shown in the inset of Fig.\ref{DWL}. The inset shows the exponent $\alpha$ in $J_1(N) \sim N^{-\alpha}$ where $N$ is varied between 128 and 256. Thus it becomes difficult to find the asymptotic exponent of thermal conductivity at low temperatures unambiguously. We are not able to get normal heat conduction in the 1D DW model, but find that asymptotic length dependence of heat current varies between $J(N) \sim 1/N^{0.79}$ to $J(N) \sim 1/N^{0.86}$ at different mean temperatures. 
\begin{figure}[htb]
\includegraphics[width=8.5cm]{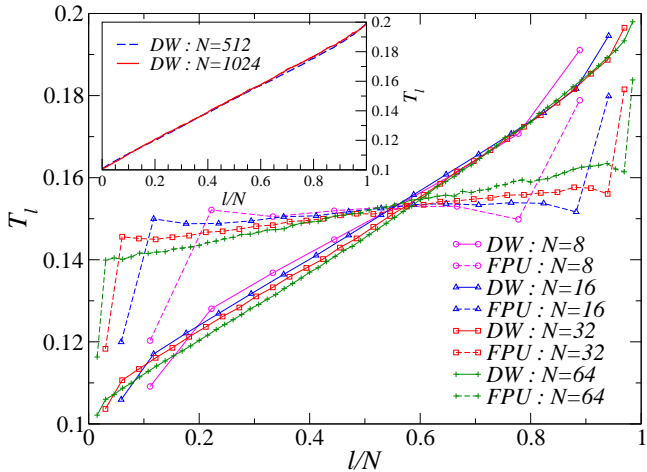}
\caption{Plot of local temperature $T_l$ with scaled site $l/N$ for the 1D DW model ($k_2=-1$, $k_4=1$) and the FPU chain ($k_2=k_4=1$) for different chain length $N$ at low temperatures. The inset shows local temperature profiles of the 1D DW model for large $N$. In all plots, $\g_{\rm L}=\g_{\rm R}=1.$} 
\label{LT}
\end{figure}
We plot local temperature $T_l$ with site $l$ of the 1D DW model for different lengths, and compare them with that of the FPU chain at low temperature (see Fig.\ref{LT}). While the 1D DW model shows local equilibration even for smallest length considered, there are large jumps in $T_l$ at the boundaries of the FPU chain for smaller lengths, and the jumps become smaller with increasing length. It confirms that it easy to get local thermalization for the DW inter-atomic potential compared to the FPU type potential. We show local temperature profiles of the DW model for longer lengths in the inset of Fig.\ref{LT}. They show a nice linear profile across the sample which signals normal heat conduction. Thus, though we could not find length independent $\kappa$, it seems that heat conduction is normal for 1D DW model at low temperature as it was concluded in Ref.\cite{Giardina00}. 

Next we study the DW model in 2D and 3D. We consider a d-dimensional hyper-cubic lattice with the DW-type inter-atomic potential. Here we denote the lattice points by the vector ${\bf n}=\{n_1,n_2,...,n_d\}$ with $n_1=1,2,...,N$, $n_{\nu=2,..d}=1,2,...,W$. The scalar displacement of an unit mass particle at the lattice site ${\bf n}$ is given by $x_{\bf n}$. The Hamiltonian of the lattice is given by
\bea
H_{d}=\sum_{\bf n}\f{1}{2}\dot{x}^2_{\bf n}+\sum_{{\bf n},\hat{\bf e}}V_{\rm DW}(x_{\bf n}-x_{{\bf n}+\hat{\bf e}})~,
\eea
where $\hat{\bf e}$ denotes the $2d$ nearest neighbors of any site. We couple all the particles at layers $n_1=1$ and $n_1=N$ to Langevin heat baths, at temperatures $T_{\rm L}$ and $T_{\rm R}$ respectively. We again use fixed boundary conditions along the direction of transport, namely the 1 direction, and employ periodic boundary conditions in the other (d-1) directions. We write Langevin equations of motion for the particles coupled to the baths, these are similar to Eqs.\ref{LE}. The corresponding white noise terms are related by fluctuation dissipation relations in Eq.\ref{FD}. We then simulate these equations using the velocity-Verlet scheme  and determine the steady-state heat current and temperature profile. The time averaged local heat current along the $1$ direction is defined following Eq.\ref{HC}. We take average of local heat current over each layer to get average heat current through a layer with constant $n_1$ consisting of $W'=W^{d-1}$ particles. In the steady state the average heat current through each layer is equal, and stationarity can be checked by testing how accurately this equality is achieved. Finally we take average of heat current over all layers to find the steady-state heat current $J_d(N)$ in the d-dimensional lattice \cite{Saito10, Chaudhuri10}.  
\begin{figure}
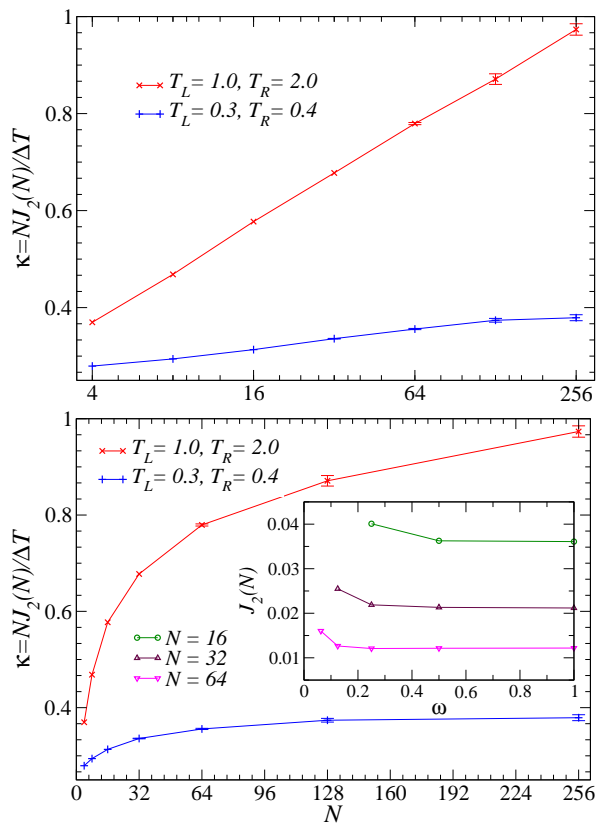

\begin{tabular}{cc}
\epsfig{file=ModelA2DJNP.eps,width=0.9\linewidth,clip=}\\
\epsfig{file=ModelA2DJN.eps,width=0.9\linewidth,clip=} 
\end{tabular}
\caption{$\kappa=NJ_2(N)/\Delta T$ with $N$ on (a) top: linear-log plot (b) bottom: double linear  plot for the 2D DW model at high and low temperatures with $k_2=-1$, $k_4=1$.   The inset in the bottom shows $J_2(N)$ with the normalized width $\omega=W/N$ for  $T_{\rm L}=1$, $T_{\rm R}=2$. In all plots, $\g_{\rm L}=\g_{\rm R}=1.$}
\label{2D} 
\end{figure}

Recently the dependence of heat current $J_d(N)$ on the width $W$ of the FPU lattice has been studied \cite{Grassberger02, Saito10}. It has been shown that for a fixed length $N$, the heat current $J_d(N)$ falls with increasing $W$ but saturates quickly to the true higher dimensional heat current. The crossover width $W_c$ at which value a crossover from 1D to higher dimensional behaviour takes place, increases slowly with $N$. We plot $J_2(N)$ for the 2D DW model with $\omega=W/N$ for different values of $N$ in the inset of Fig.\ref{2D}. We find that the crossover from 1D to 2D nature occurs at decreasing $\omega$ for increasing $N$.  It indicates that we can calculate heat current accurately in higher dimensional longer lattices by simulating relatively smaller $W$. We employ this observation in our simulations for longer lattices in both 2D and 3D.     

Next we check the dependence of heat current $J_2(N)$ on $N$ for the 2D DW lattice  at high and low temperatures. An estimate of the crossover temperature $\bar{T}_d$ separating two different structural phases in higher dimensions is more complicated than 1D, but we can roughly estimate it as $\bar{T}_d \sim d~\bar{T}_1$ for a d-dimensional homogeneous lattice. The behaviour of heat current at high temperature of the baths, $T_{\rm L}=1$ and $T_{\rm R}=2$ is shown in the top of Fig.\ref{2D} using linear-log scale which clearly shows $\kappa \sim {\rm log}(N)$. Thus we confirm the predicted logarithmic behaviour of 2D FPU lattices in our simulations for the DW model with relatively shorter lengths. It is interesting to mention that this length dependence of $\kappa$ in 2D DW model at high temperature is different from the previous simulated behaviour of 2D FPU lattices. The variation of the temperature gradient is nonmonotonic as a function of distance across the lattice in 2D at high temperatures (see Fig.\ref{LT3d}) which indicates non-diffusive transport as argued in Ref.\cite{Saito10}.  The length dependence of steady state heat current $J_2(N)$  at low temperature is nonmonotonic with increasing mean temperature which is similar to the 1D case. We find clear signature of normal heat conduction in the 2D DW model at low temperature as shown in the bottom of Fig.\ref{2D} using double linear scale. The corresponding  local temperature profile is fully linear. 

\begin{figure}[htb]
\includegraphics[width=8.5cm]{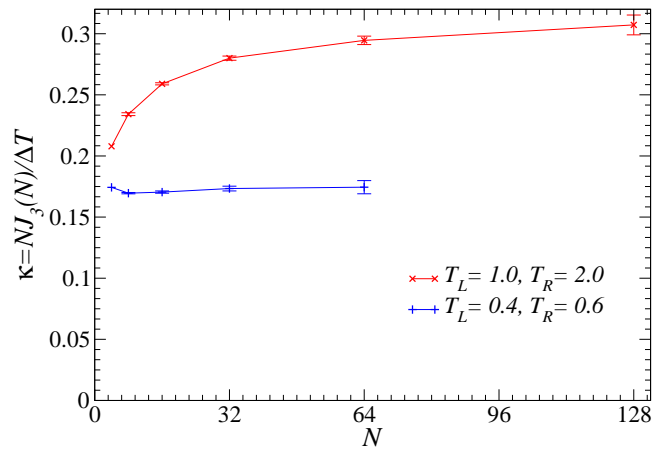}
\caption{$\kappa=NJ_3(N)/\Delta T$ with $N$ for the 3D DW model at high and low temperatures with $k_2=-1$, $k_4=1$. In all plots, $\g_{\rm L}=\g_{\rm R}=1.$}
\label{3D} 
\end{figure}

Finally we consider behaviour of $J_3(N)$ for the 3D DW model. We readily find normal diffusive heat conduction at low temperature for quite shorter lengths. This is shown in Fig.\ref{3D} using double linear  plot for bath temperatures $T_{\rm L}=0.4$ and $T_{\rm R}=0.6$. The corresponding local temperature profile is also linear as shown in the inset of Fig.\ref{LT3d}. The error bars  in our simulated heat current for longer lengths are significant (about 2.5$\%$ for N=128), thus we could not confirm unambiguously normal heat conduction for the 3D DW model at high temperature.  Though we find signature of normal heat conduction showing length independent $\kappa$ at high temperatures (shown in Fig.\ref{3D} for $T_{\rm L}=1.0$ and $T_{\rm R}=2.0$).  We strongly believe that heat conduction in 3D at high temperature for the DW model is normal and simulation for longer lengths would confirm it (even for much smaller length compared to Ref.\cite{Saito10}). We justify our claim for normal heat conduction for the 3D DW model by comparing local temperature profiles of 1D, 2D and 3D DW models at high temperature. It is shown in  Fig.\ref{LT3d} which shows that the variation of the temperature gradients in both 1D and 2D are nonmonotonic as a function of distance across the lattice while for the 3D DW model it is almost monotonic and linear which are characteristics of normal diffusive behaviour \cite{Saito10}.
\begin{figure}[htb]
\includegraphics[width=8.5cm]{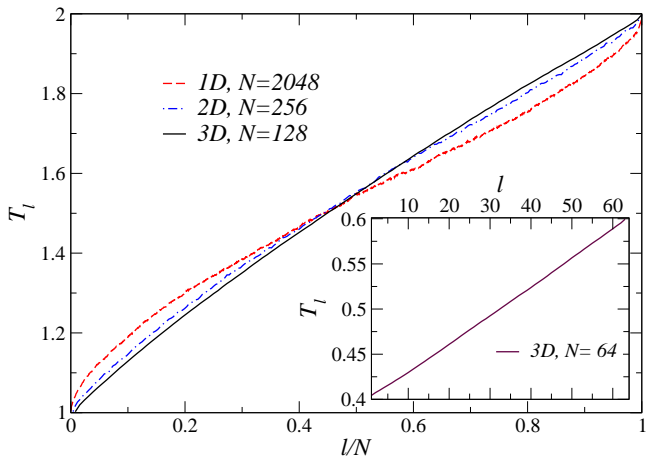}
\caption{The plot of local temperature of 1D, 2D, 3D DW model for $T_{\rm L}=1$, $T_{\rm R}=2$. The inset shows local temperature for 3D DW model for $T_{\rm L}=0.4$, $T_{\rm R}=0.6$. In all plots, $\g_{\rm L}=\g_{\rm R}=1.$} 
\label{LT3d}
\end{figure}

A dimensional crossover of thermal transport is recently observed in few-layer graphene \cite{Ghosh10}. The type of crossover in thermal transport discussed here can be realized for the case of displacement ferroelectric transitions in which a crystal undergoes a structural transition from high temperature paraelectric state with higher symmetry to a low temperature ferroelectric state with a barrier at the normal lattice site \cite{Landau}.

The author is indebted to  Prof. A. J. Sievers for valuable discussions.

\end{document}